\documentclass[preprint,prc,superscriptaddress,unsortedaddress,showpacs,preprintnumbers,amsmath,amssymb]{revtex4}
\usepackage{tipa}

\usepackage{graphicx}
\usepackage{dcolumn}
\usepackage{bm}

\def\beq{\begin{equation}}
\def\eeq{\end{equation}}
\def\bea{\begin{eqnarray}}
\def\eea{\end{eqnarray}}

\def\fun#1#2{\lower3.6pt\vbox{\baselineskip0pt\lineskip.9pt
  \ialign{$\mathsurround=0pt#1\hfil##\hfil$\crcr#2\crcr\sim\crcr}}}

\preprint{}

\begin{document}

\title{Three-body force effect on off-shell mass operator and spectral functions in nuclear matter}

\author{ Pei Wang}
 \affiliation{Institute
of Modern Physics, Chinese Academy of Sciences, Lanzhou 730000,
China} \affiliation{University of Chinese Academy of Sciences,
Beijing 100049, China}
\author{ Sheng-Xin Gan }
 \affiliation{Institute
of Modern Physics, Chinese Academy of Sciences, Lanzhou 730000,
China} \affiliation{University of Chinese Academy of Sciences,
Beijing 100049, China}
\author{ Peng Yin}
 \affiliation{Institute
of Modern Physics, Chinese Academy of Sciences, Lanzhou 730000,
China} \affiliation{University of Chinese Academy of Sciences,
Beijing 100049, China}
\author{Wei Zuo}\email[ ]{zuowei@impcas.ac.cn}
\affiliation{Institute of Modern Physics, Chinese Academy of
Sciences, Lanzhou 730000, China}

\begin{abstract}
Within the framework of the Brueckner theory, the off-shell behaviors
of the mass operator $M(k,\omega)=V(k,\omega)+iW(k,\omega)$, i.e.,
its dependence upon the momentum $k$ and upon the nucleon
frequency $\omega$, are investigated by including
nuclear three-body force (TBF). The first two terms of the
hole-line expansion of the mass operator are taken into account.
The TBF effects on their off-shell properties are discussed. A
comparison is made between the on-shell and off-shell values of
$M_{1}$. The nucleon spectral function and nucleon momentum
distribution are also calculated, and the calculation shows that
they are hardly affected by the TBF effect at the saturation
density. At a high density two times greater than the saturation density,
inclusion of the TBF may lead to a visible effect on the spectral
function and may enhance the depletion of the hole states.

\end{abstract}
\pacs{21.65.-f, 21.30.Fe, 24.10.Cn, 21.10.Pc}

\maketitle

\section{Introduction}
The probability of removing a particle with momentum $k$ from a
target nuclear system, leaving the final system with excitation
energy $\omega$, is reflected by the nucleon spectral function
$S(k,\omega)$. In the free Fermi gas model, the spectral function
can be written as $S(k,\omega)=\delta(\omega-\hbar^{2}k^{2}/2m)$.
However, many-body correlations among the nucleons, induced by the
nucleon-nucleon ($NN$) interactions, broaden the peaks of the
Fermi-gas spectral function and decrease their strengths. It has
been shown that, most of this decrease of the strength of the
single-particle (s.p.) states with respect to the standard mean
field estimates~\cite{1} is due to the $NN$
correlations~\cite{2,3}, whose effects can be most accurately
studied in infinite nuclear matter. Therefore, a microscopic
calculation of the nucleon spectral function in nuclear matter is
of special interest since it may play an important role in
understanding the nature of the $NN$ correlations, especially the
short-range and tensor
correlations~\cite{pandharipande:1997,dickhoff:2004}.

The interest in the spectral function has also been
raised by the treatment of the off-shell effect in transport theory,
which provides a generalized theoretical framework to describe the time
evolution of heavy ion reactions.
The use of the quasiparticle
approximation (QPA) in transport theory puts the nucleon on the mass
shell, neglecting not only the finite decay width of the particles,
but also the width of the nucleon spectral function. However, the
on-shell quasiparticle limit should not be adequate for particles
with short lifetimes and/or high collision rates as recognized previously \cite{5,6}.
Therefore, there have been attempts to go beyond the
QPA, such as transport formulations for quasiparticles with
dynamical spectral functions \cite{7,8,9}, and extend the
extensively applied models \cite{10,11,12,13,14,15,16,17,18,19,20}.
The nucleon spectral function plays an important role in the
implementation of the off-shell effects in a transport theoretical
treatment of heavy-ion and other nuclear collisions.

Experimentally, the information about the nuclear spectral function and the effect of $NN$ correlations in
nuclear systems can be extracted from
the ($e,e'p$) and proton-induced knockout
reactions \cite{pandharipande:1997,dickhoff:2004,dickhoff:1992,mitt:1990,lapikas:1999,starink:2000,batenburg:2001}.
Theoretically,  the nuclear short-range correlations and the spectral function in nuclear matter
have been investigated extensively by using various microscopic nuclear many-body approaches, such as the Green function
 theory~\cite{vonderfecht:1991,vonderfecht:1993,ramos:1989,muther:1995,dewulf:2002,Dieperink:2003,frick:2005,rios:2009},
 the correlated basis function method~\cite{fantoni:1984,benhar:1989,benhar:1990,benhar:1992},
 the extended Brueckner-Hartree-Fock (EBHF)  framework~\cite{baldo:1990,baldo:1991,4,frick:2002,hassaneen:2004}, and
 the in-medium $T$-matrix approach~\cite{bozek:1999,bozek:2002,soma:2008}. For a review, we refer readers to
 Refs.~\cite{pandharipande:1997,dickhoff:2004}.
 Within the framework of the Brueckner theory, the nucleon spectral function in symmetric nuclear matter has
 been studied in Ref.~\cite{4} by adopting a finite-rank representation of the realistic Argonne $V14$ $NN$ interaction,
 without taking into account any three-body force (TBF) effect. In Refs.~\cite{hassaneen:2004,rios:2009},
 the neutron and proton spectral functions in isospin-asymmetric
  nuclear matter have been explored using the BHF approach and the Green function theory, respectively.
  It is well known that inclusion of TBF in the nonrelativistic Brueckner theory is crucial for reproducing
 the nuclear saturation properties and for better describing the s.p. properties, such as the momentum dependence of the
 nucleon s.p. potential~\cite{26,27,zuo:2005,baldo}.
 Recently, the TBF effect on the spectral function in nuclear matter has been investigated within the framework of
 the in-medium $T$-matrix method in Ref.~\cite{soma:2008}, where the TBF adopted is the Urbana TBF~\cite{carlson}.
In that paper, the authors have shown that the TBF effect on the spectral functions is quite small
at low densities around and below the saturation density and that noticeable modification of the spectral functions
is realized only for high densities well above the saturation density.
One of our purposes in the present paper is to investigate the possible
impact of a microscopic TBF on the nucleon spectral function
within the framework of the extended BHF approach.

The spectral function is closely related to the mass operator
$M(k,\omega)$, whose off-shell behavior is also our concern in
the present paper. The off-shell mass operator plays an important role in
the dispersion relation to the nuclear mean
field~\cite{21} and in the discussion of $y$-scaling in
inclusive electron scattering \cite{22,23,24}. In Ref. \cite{4},
the properties of the off-shell mass operator have been obtained
within the Brueckner theory in the absence of any TBF. Therefore,
the other purpose of the present paper is to reveal the TBF effect on the
off-shell mass operator discussed in Ref.~\cite{4}.

The paper is organized as follows. In Sec. II, we give a brief review of the mass operator
and spectral function, i.e., their definitions and physical interpretations, as presented in Ref. \cite{4}.
We also provide a simple introduction to the Brueckner theory and the microscopic TBF adopted in our
calculation. In Secs. III and IV, we focus on the real and imaginary parts of the off-shell mass
operator. We study the $k$-dependence
of $M(k,\omega)$ for the two typical energies, $\omega=20$ MeV and $\omega=160$ MeV, at 0.34 fm$^{-3}$.
We also calculate the $\omega$-dependence of $M(k,\omega)$ for $k={3 \over 4}k_{F}$ and $k={5 \over 4}k_{F}$. The off-shell results
of $M_{1}(k,\omega)$ are compared with the on-shell ones.
The TBF effect on the $k$ and $\omega$-dependence of $M(k,\omega)$ is discussed.
In Sec. V, we calculate the spectral function and investigate the TBF effect
on its $\omega$-dependence. In Sec. VI, a summary is given.

\section{formalism}
\subsection{The mass operator and the spectral function}
The Green function in the energy-momentum representation is given by
$G(k,\omega)=[\omega-k^{2}/2m-M(k,\omega)]^{-1}$, where $M(k,\omega)=V(k,\omega)+iW(k,\omega)$
is the mass operator that can be identified with the mean field felt by a nucleon in a nuclear system.
The real and imaginary parts of the mass operator are connected by the dispersion relation \cite{4}:
\begin{equation}
V(k,\omega)=\lim _{\omega\to \infty}V(k,\omega)+{1 \over \pi}\int _{-\infty}^{\infty}{W(k,\omega') \over \omega'-\omega}\,d\omega' \ .
\end{equation}
The spectral function is given by
\begin{equation}\label{eq:S}
S(k,\omega)=-{1 \over \pi}{W(k,\omega)\overwithdelims () [\omega-k^{2}/2m-V(k,\omega)]^2+[W(k,\omega)]^2} \ ,
\end{equation}
and it fulfills the sum rule
\begin{equation}
\int _{-\infty}^{\infty}S(k,\omega)\,d\omega=1.
\end{equation}
The occupation probability $n(k)$ is related to the spectral function by
\begin{equation}\label{eq:n1}
n(k)=\int _{-\infty}^{\omega_{F}}S(k,\omega)\,d\omega
\end{equation}
and
\begin{equation}\label{eq:n2}
n(k)=1-\int _{\omega_{F}}^{\infty}S(k,\omega)\,d\omega \ .
\end{equation}
The Fermi energy $\omega_{F}$ fulfills $\omega_{F}=k^{2}_{F}/2m+V(k_{F},\omega_{F})$.
For a system of $A$ nucleons, $S(k,E^*)$ measures the probability density of finding the
residual ($A-1$)-nucleon system with excitation energy $E^*=\omega_{F}-\omega (\omega<\omega_{F})$
after removing a nucleon with momentum $k$ from the ground state, or the probability density
of finding the residual ($A+1$)-nucleon system with the excitation energy
$E^*=\omega-\omega_{F} (\omega>\omega_{F})$ after one has added a nucleon
with momentum $k$ to the ground state.

\subsection{Brueckner theory with a microscopic TBF}
The starting point of Brueckner calculation of nuclear matter properties is to obtain
the Brueckner reaction matrix $G(\omega)$ by solving the Bethe-Goldstone (BG) equation
\begin{equation}
G(\omega)=V_{\rm NN}+V_{\rm NN}\sum \limits_{k_1k_2}{{|k_1k_2 \rangle Q(k_1k_2)
\langle k_1k_2|} \over {\omega - \epsilon (k_1)-\epsilon(k_2)+i\eta}}G(\omega) \ ,
\end{equation}
where $k_1$ and $k_2$ are momenta of the two involved nucleons. $Q (k_1,
k_2)=[1-n(k_1)][1 -n(k_2)]$ is the Pauli operator which prevents
two intermediate nucleons from being scattered into occupied
states. $\omega$ is the starting energy. The single-particle
energy $\epsilon(k)$ satisfies the on-shell relation
$\epsilon(k)=k^2/2m+U_{\rm BHF}(k)$, where the auxiliary
potential $U_{\rm BHF}(k)$ is the single-particle potential at the
BHF level and it is defined as $U_{\rm
BHF}(k)=\sum \limits_{k'}^{}{\rm Re}\langle
kk'|G(\epsilon(k)+\epsilon(k'))|kk' \rangle_A $. The subscript $A$
denotes antisymmetrization of the matrix element. The continuous
choice other than gap choice is adopted when solving the BG
equation to obtain the $G$-matrix \cite{25}.

Extension of the Brueckner-Bethe-Goldstone (BBG) theory to include
TBFs can be found in Refs.~\cite{26,27}. In this paper, we choose
the microscopic TBF which is based on the meson-exchange current
model proposed by P. Grang\'{e} {\it et al.} \cite{28} and reduced
to an equivalent effective two-body force $V_{3}^{\rm eff}$ via an
average with respect to the third-nucleon degree of freedom.
The effective force $V_{3}^{\rm eff}$ in $r$ space reads
\begin{align}\label{eq:V3eff}
V_{3}^{\rm eff}({\vec r}'_1, {\vec r}_2 '|{\vec r}_1, {\vec r}_2)
&={1 \over 4}Tr\sum \limits_{n}\int d\,{\vec r}_3 d\,{\vec r}_3' \phi_n^\ast({\vec r}_3')[1-\eta(r_{13}')]\nonumber\\
&\times [1-\eta(r_{23}')]W_3({\vec r}_1', {\vec r}_2 ',{\vec r}_3' |{\vec r}_1, {\vec r}_2,{\vec r}_3)\nonumber\\
&\times \phi_n({\vec r}_3)[1-\eta(r_{13})][1-\eta(r_{23})] \ ,
\end{align}
where the wave function $\phi_n$ denotes the single nucleon wave function in free space.
The realistic $NN$ interaction $V_{NN}$ in the BG equation is the
sum of the Argonne $V_{18}$ ($AV18$) two-body interaction and the
effective two-body force $V_{3}^{\rm eff}$, as described in
Refs.~\cite{26,27}. Since $\eta(r)$ in expression
({\ref{eq:V3eff}) is the so-called defect function \cite{28,29}
corresponding to the $G$-matrix, $V^{\rm eff}_3$ should be
recalculated along with the $G$-matrix in each iteration of our
BHF procedure to ensure self-consistency of the BG equation.

In the spirit of Brueckner theory, the first two terms of the hole-line expansion
of the mass operator are the BHF approximation $M_{1}(k,\omega)$ and the Pauli rearrangement correction
$M_{2}(k,\omega)$. They are represented by the diagrams of Fig.1, and their expressions read :
\begin{equation}
M_{1}(k,\omega)=\sum_{h<k_{F}}\langle kh \vert G[\omega+\epsilon(h)] \vert kh \rangle_{A},
\end{equation}
\begin{equation}
M_{2}(k,\omega)={1 \over 2}\sum_{l,m<k_{F},n>k_{F}}{\vert \langle lm \vert G[\epsilon(l)+\epsilon(m)] \vert kn \rangle_{A} \vert^{2} \over \omega+\epsilon(n)-\epsilon(l)-\epsilon(m)-i\delta} .
\end{equation}
Their off-shell values can be calculated as long as the $G$-matrix
is obtained.

\begin{figure}[htbp]
\begin{center}
\includegraphics[width=7cm]{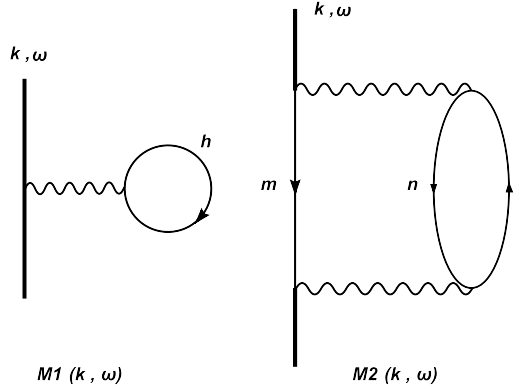}
\caption{\label{fig1} Representation of the first two terms of the
hole-line expansion of the mass operator. The thin lines represent
either particle (upward-pointing arrows) or hole (downward-pointing arrows)
momentum states. The thick lines show the
values of the nucleon momentum $k$ and frequency $\omega$.}
\end{center}
\end{figure}

\section{Frequency dependence of the mass operator at fixed momentum}
\subsection{Real part of the off-shell mass operator}
The calculated $\omega$-dependence of
$V_{1}(k,\omega)={\rm Re}M_{1}(k,\omega)$ and
$V_{2}(k,\omega)={\rm Re}M_{2}(k,\omega)$ is shown in Fig.~\ref{fig2}
for the two densities of $\rho=0.17$ fm$^{-3}$ and $\rho=0.34$
fm$^{-3}$, respectively. Two fixed momenta ($k={3 \over 4}k_{F}$
and $k={5 \over 4}k_{F}$) are selected. The quantity $e_{F}$ is
the calculated value of the single-particle energy $\epsilon(k)$
at the Fermi momentum: $e_{F}=\epsilon(k_{F})=k_{F}^2/2m + U_{\rm
BHF}(k_{F})$. As one can see in Fig.~\ref{fig2}, the quantity
$V_{1}(k,\omega)$ is attractive for $\omega<e_{F}$ and its
attraction increases as a function of frequency $\omega$ in the
region of $\omega<e_{F}$, while $V_{2}(k,\omega)$ is repulsive for
$\omega>e_{F}$ and its repulsion decreases with increasing
$\omega$ in the region of $\omega>e_{F}$. The TBF effect on their
$\omega$-dependence is also reported in this figure. Inclusion of
the TBF in our calculations hardly affects the $\omega$-dependence
of $V_{2}(k,\omega)$, but tends to reduce the attraction of
$V_{1}(k,\omega)$ well below $e_{F}$ and enhance its attraction as
$\omega$ is much larger than $e_{F}$. At the saturation density of
0.17 fm$^{-3}$, the TBF effect on $V_{1}(k,\omega)$ is weak enough
to be neglected in the vicinity of $e_{F}$. However, the TBF
effect gets much stronger at high densities. As a result, the
TBF-induced reduction of the attraction of $V_{1}(k,\omega)$ well
below $e_{F}$ is obviously seen at two times the saturation
density 0.34 fm$^{-3}$, as revealed in the right panel of
Fig.~\ref{fig2}. At high densities, the TBF effect on $V_2$
turns out to be rather small. At $\rho=0.34$fm$^{-3}$ and
$k=2.1$fm$^{-1}$, inclusion of the TBF may enhance slightly the
repulsion of $V_2$.

Besides, it is worth noticing that the distinct deviation of
the curve with open squares from that with filled squares when $\omega-e_{F}$ is above 150 MeV.
The deviation appears regardless of the density value, which indicates that
one should account for the TBF effect carefully in the high-energy domain.

\subsection{Imaginary part of the off-shell mass operator}

Figure \ref{fig3} shows the dependence of $W_{1}(k,\omega)$ and $W_{2}(k,\omega)$ upon the
difference $\omega-e_{F}$. One important feature of the two components is
that $W_{1}(k,\omega)$ vanishes for $\omega<e_{F}$ and $W_{2}(k,\omega)$ vanishes
for $\omega>e_{F}$. Moreover, $W_{2}(k,\omega)$ also vanishes for large
negative $\omega$.
At the saturation density of 0.17 fm$^{-3}$, the calculated $W_{1}(k,\omega)$
including the TBF contribution is very close to its values without including the TBF
contribution in the energy domain ranging from $e_{F}$ to approximately 300 MeV.
However, as density increases to 0.34 fm$^{-3}$ where the TBF effect
becomes strong, inclusion of the TBF leads to a faster increase of the attraction
of $W_{1}(k,\omega)$ with increasing frequency $\omega$ as
compared to the result without the TBF effect. At $\rho=0.34$fm$^{-3}$ and $k=2.1$fm$^{-1}$,
inclusion of the TBF may lead to a sizable enhancement of the attraction of $W_2$.

\subsection{Comparison with on-shell values}
In Fig.~\ref{fig4}, we compare the off-shell values of $W_{1}(k,\omega)$ and
$W_{2}(k,\omega)$ with their on-shell values.
Although our calculations are done at a higher density of 0.34 fm$^{-3}$ and
in the presence of the TBF, the results plotted in Fig.~\ref{fig4}
are similar qualitatively to those in Fig.~12 of Ref. \cite{4}, regardless of the magnitude.
Therefore, the analysis and conclusion in Ref. \cite{4} remain valid.
That is to say, on the one hand, $W_{1}(k,\omega)$ and $W_{2}(k,\omega)$ are symmetric with each other
only in the vicinity of the Fermi energy; on the other hand, the assumption in the simplest version of the
dispersion relation approach for the nuclear mean field [ i.e., the $\omega$-dependence of $W_{1}(k,\omega)$
is approximated by the $e$-dependence of the on-shell $W_{1}(e)$]  is only justified qualitatively.

\begin{figure*}[htbp]
\begin{center}
\includegraphics[width=.9\textwidth,clip]{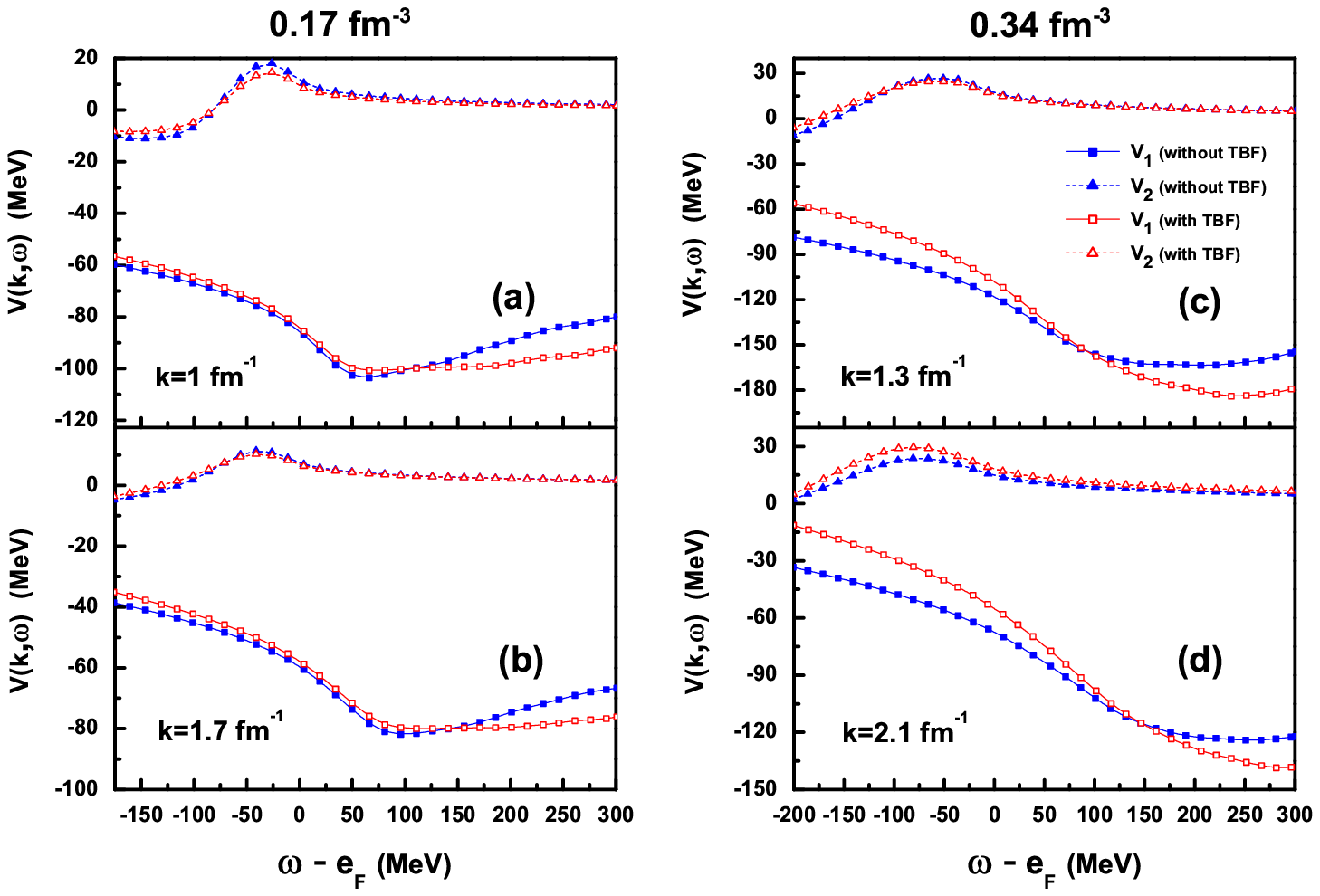}
\caption{\label{fig2} (Color online) Dependence of $V_{1}(k,\omega)$ and
$V_{2}(k,\omega)$ upon $\omega-e_{F}$ for the two densities of
$\rho=0.17$ fm$^{-3}$ and $\rho=0.34$ fm$^{-3}$, and for the two fixed
momenta of $k={3 \over 4}k_{F}$ and $k={5 \over 4}k_{F}$. The
curves with open squares and open triangles have taken into
account the TBF contribution.}
\includegraphics[width=.9\textwidth,clip]{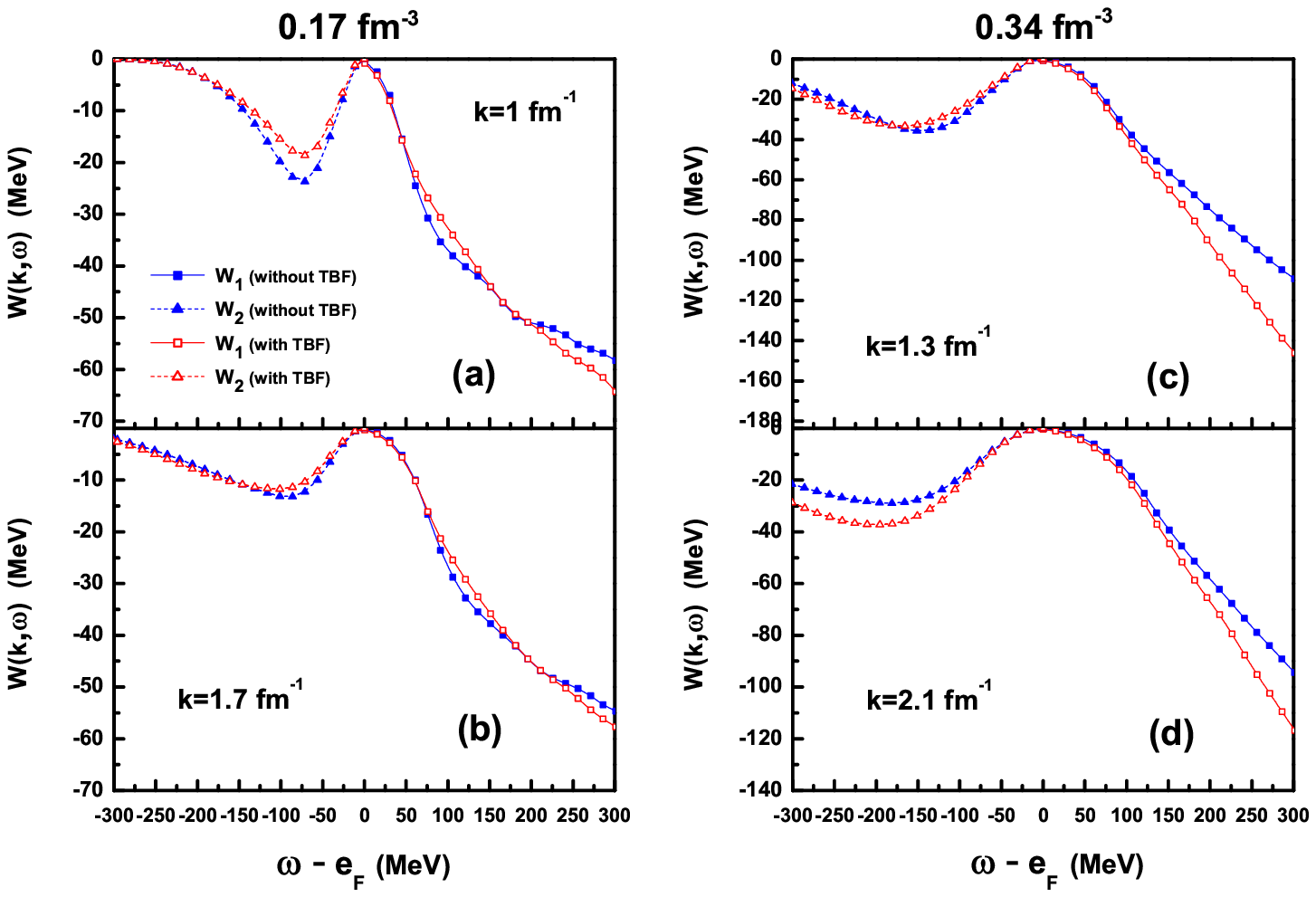}
\caption{\label{fig3} (Color online) Dependence of $W_{1}(k,\omega)$ and
$W_{2}(k,\omega)$ upon $\omega-e_{F}$ for the two densities of
$\rho=0.17$ fm$^{-3}$ and $\rho=0.34$ fm$^{-3}$, and for the two fixed
momenta of $k={3 \over 4}k_{F}$ and $k={5 \over 4}k_{F}$. The
curves with open squares and open triangles have taken into
account the TBF contribution.}
\end{center}\end{figure*}
\begin{figure}[htbp]
\begin{center}
\includegraphics[width=9cm]{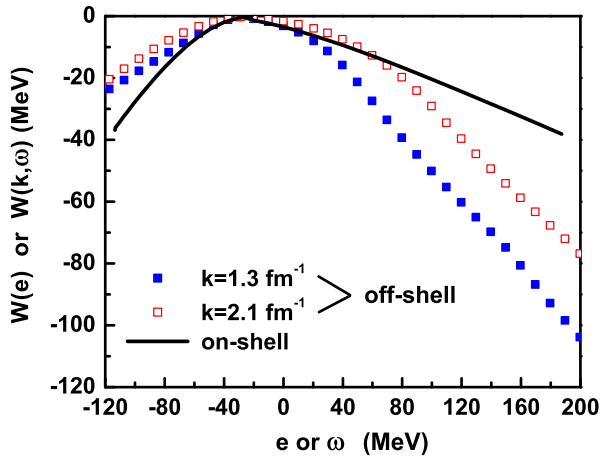}
\caption{\label{fig4} (Color online) Comparison between the $e$-dependence of the
on-shell $W_{1}(e)$ and $W_{2}(e)$ (solid curve) with the
$\omega$-dependence of $W_{1}(k,\omega)$ and of $W_{2}(k,\omega)$,
for $k=1.3$ fm$^{-1}$ (filled squares) and $k=2.1$ fm$^{-1}$ (open
squares), at the density of 0.34 fm$^{-3}$.}
\end{center}\end{figure}

\section{Momentum dependence of the BHF field $M_{1}(k,\omega)$ at fixed frequency}

At the density of 0.34 fm$^{-3}$, we calculate the dependence of
$V_{1}(k,\omega)$ and $W_{1}(k,\omega)$ upon momentum $k$ for two
fixed frequencies, namely $\omega_{1}=20$ MeV and $\omega_{2}=160$
MeV. The corresponding on-shell values of the momentum, which can
be obtain from the energy-momentum relation
$\omega(k)=k^{2}/2m+V[k,\omega(k)]$, are approximately
$k(\omega_{1})=2.15$ fm$^{-1}$ and $k(\omega_{2})=3.15$ fm$^{-1}$.
Results for the saturation density 0.17 fm$^{-3}$ are not
provided, because the TBF effect is not expected to play an
important role at such a relatively low density.

In Fig.~\ref{fig5}, the upper part displays the calculated values of $V_{1}(k,\omega=20 {\rm MeV})$ and
$V_{1}(k,\omega=160 {\rm MeV})$;
the lower part presents the calculated values of $W_{1}(k,\omega=20 {\rm MeV})$ and
$W_{1}(k,\omega=160 {\rm MeV})$.

As we can see from the figure, for both the real and the imaginary parts of $M_{1}(k,\omega)$, the open squares
are very close to the corresponding filled squares in the high momentum region, indicating that the TBF correction is
small at high momenta.
However, in the low momentum region, the TBF has a strong effect on the shape of the $k$-dependence of
$V_{1}(k,\omega)$ and $W_{1}(k,\omega)$ only at the larger frequency $\omega=160$ MeV, and
it may separate the open squares and the
corresponding filled ones considerably. As a result, it is necessary
to take into account the TBF effect if one wants to get more exact and reliable
$k$-dependence of the off-shell mean field $M_{1}(k,\omega)$ felt by a nucleon with both low momentum and large frequency.

\section{Spectral function}

The spectral function $S(k,\omega)$ can be calculated from
Eq.(\ref{eq:S}), using the real and imaginary parts of the mass
operator. Notice that in the present approximation scheme,
$W(k,\omega)=W_{2}(k,\omega)$ for $\omega<e_{F}$ and
$W(k,\omega)=W_{1}(k,\omega)$ for $\omega>e_{F}$. For energies
$\omega<e_{F}$, the spectral function $S(k,\omega)$ is referred to
as the ``hole spectral function" $S_{h}(k,\omega)$, and for
energies $\omega>e_{F}$, the $S(k,\omega)$ becomes the
``particle spectral function" $S_{p} (k,\omega)$.
$S_{h(p)}(k,\omega)$ measures the probability that a nucleon with
momentum $k$ and energy $\omega$ can be removed from (added to)
the ground state.

In Fig.~\ref{fig6}, the spectral function is plotted versus
$\omega$ at the density of 0.34 fm$^{-3}$. The upper part of the
figure displays the spectral distribution for momentum below the
Fermi momentum. In the independent-particle model, states with
momenta below the Fermi surface would be completely occupied so
that the spectral function is identical to a $\delta$ function
located at the on-shell value of $\omega$. However, the two-hole
configuration leads to a non vanishing imaginary part of the mass
operator and consequently a finite spectral function peaked at the
on-shell energy for momenta below
$k_{F}$~\cite{pandharipande:1997,dickhoff:2004}. The quasiparticle
peak in the spectral function can be related to the shell model by
the fact that when a nucleon with momentum $k$ is removed from
\begin{figure}[htbp]\begin{center}
\includegraphics[width=7cm]{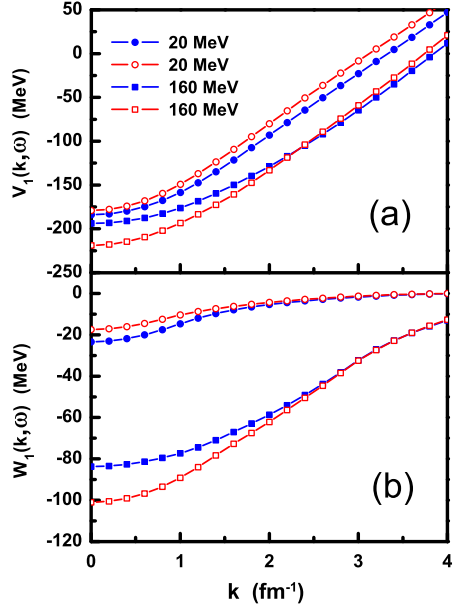}
\caption{\label{fig5} (Color online) Dependence upon $k$ of the calculated values
of $V_{1}(k,\omega)$ and of $W_{1}(k,\omega)$ for the two selected
frequencies: $\omega=20$ MeV and $\omega=160$ MeV. Open symbols
correspond to the case with TBF contribution, while filled symbols
do not. The density is fixed at 0.34 fm$^{-3}$.}
\end{center}\end{figure}
\begin{figure}[htbp]\begin{center}
\includegraphics[width=7cm]{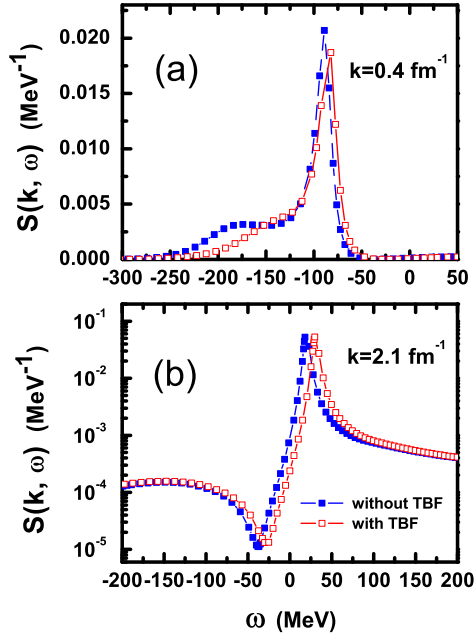}
\caption{\label{fig6} (Color online) Spectral function $S(k,\omega)$ calculated
from Eq.(\ref{eq:S}) at the density of 0.34 fm$^{-3}$.}
\end{center}\end{figure}
the ground state, the residual system has a large probability of
having a well-defined excitation energy $E^{*}_{A-1}$ \cite{4}. The
lower part of the figure shows the spectral distribution for
momentum above $k_{F}$.

Recently, the TBF effect on the spectral function in nuclear matter has been investigated explicitly
within the in-medium $T$-matrix method in Ref.~\cite{soma:2008} where the
Urbana TBF~\cite{carlson} has been adopted.
One may notice from the upper panel of Fig.~\ref{fig6} that, at
momentum below the Fermi momentum $k_F$, the TBF effect on the
spectral distribution leads to a shift of the peak location to
slightly higher energy and a decrease in the peak value, in
agreement with the results of Ref.~\cite{soma:2008} within the
in-medium $T$-matrix method using the Urbana TBF. It is
also seen that the TBF reduces the strength of the spectral
distribution at large negative energies. At momentum above
$k_F$, the TBF effect is mainly to shift the peak value to a
higher energy. The TBF-induced shift of the peak location of the
spectral distribution can be understood readily since the TBF
gives an extra repulsive contribution to the on-shell
single-particle potential and consequently increases the on-shell
energy for a given momentum $k$.

In order to test the numerical accuracy of the present work, in
Fig.~\ref{fig7} we display the nucleon momentum distribution
defined in Eq.~(\ref{eq:n1}) for two densities $\rho=0.17$ and
0.34fm$^{-3}$. By using Eq.~(\ref{eq:n2}) we get
almost the same results. Due to the nucleon-nucleon correlations,
the s.p. hole states below $k_F$ are partly empty and the particle
states above $k_F$ are partly occupied in the correlated ground
state of nuclear matter. The depletion of the lowest hole state at
$k=0$ at $\rho=0.17$fm$^{-3}$ is about $16.4\%$, which
is compatible with the previous predictions in
\begin{figure}[htbp]\begin{center}
\includegraphics[width=0.5\textwidth]{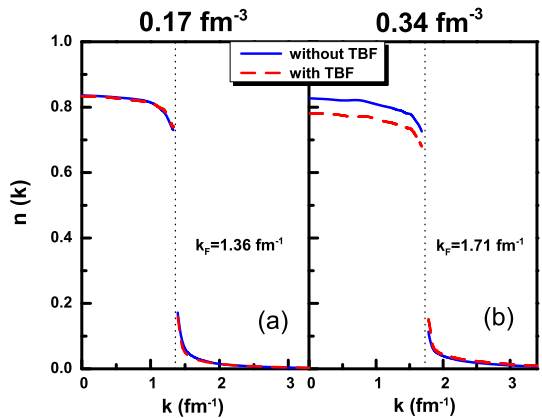}
\caption{\label{fig7} (Color online) Nucleon momentum distribution in symmetric
nuclear matter at two densities $\rho=0.17$ fm$^{-3}$ (left panel)
and $\rho=0.34$ fm$^{-3}$ (right panel)}
\end{center}\end{figure}
Refs.~\cite{dickhoff:2004,baldo:1991,fantoni:1984,benhar:1990,rios:2009,frick:2002}.
This value is also consistent with the experimental result in
Ref.~\cite{batenburg:2001}. As discussed in
Ref.~\cite{baldo:1991}, inclusion of the (higher order)
renormalization contribution $M_3$ in the mass operator may reduce
the calculated depletion from $\sim 17\%$ to $\sim 14\%$ by
using a separable $AV14$ interaction. It is noticed that the TBF
effect is negligibly small at the saturation density
$\rho=0.17$fm$^{-3}$, in agreement with the conclusion of
Ref.~\cite{fantoni:1984} within the correlated basis function
approach by adopting the Urbana $v14$ interaction plus an
effective TBF. The TBF effect only becomes sizable at high
densities well above the saturation density as shown in the right
panel of Fig.~\ref{fig7} where the momentum distribution for
$\rho=0.34$ fm$^{-3}$ is plotted. The TBF effect is shown to
enhance the depletion of the hole states since the TBF may induce
sufficiently strong extra short-range correlations at sufficiently high
densities. At $\rho=0.34$ fm$^{-3}$,
 inclusion of the TBF may enhance the depletion of
 the zero-momentum state from $\sim 17\%$ to $\sim 22\%$.

\section{summary}

Within the framework of Brueckner theory extended to include a
microscopic TBF, we have calculated the dependence of the
off-shell mass operator upon the momentum $k$ and upon the nucleon
frequency $\omega$. The first two terms in the hole-line expansion
of the mass operator are taken into account. Our calculations show
that the TBF effect on the values of $M_{1}(k,\omega)$ for fixed
momentum is only important at high densities or at frequencies far
away from its on-shell energy at $k_{F}$. However, the
$\omega$-dependence of the Pauli rearrangement term
$M_{2}(k,\omega)$ at fixed momenta is even less affected by the
TBF effect. At $\rho=0.34$ fm$^{-3}$ which is well above the
saturation density, inclusion of the TBF may enhance the repulsion
of $V_2$  at a large momentum $k=2.1$ fm$^{-1}$ above $k_F$. We
also compare the off-shell values of $M_{1}$ at fixed momenta with
its on-shell values. For fixed frequency, the $k$-dependence of
the BHF field $M_{1}$ is investigated, and it is shown that it is
necessary to take into account the TBF effect if one wants to get
a more exact $k$-dependence of the mean field $M_{1}(k,\omega)$
felt by a nucleon with both low momentum and large frequency. The
nucleon spectral function has been calculated. At density of
$\rho=0.34$ fm$^{-3}$ well above the saturation density, the TBF
effect shifts the peak location in the spectral function to
slightly higher energy and reduces slightly the peak value at
low momentum below the Fermi momentum $k_F$. The TBF effect on
the nucleon spectral function and
nucleon momentum distribution turns out to be neglected at the
saturation density $\rho=0.17$fm$^{-3}$.  It becomes sizable
only at high densities well above the saturation density, and inclusion of
the TBF leads to an enhancement of the depletion of the zero-momentum hole
state from $\sim 17\%$ to $\sim 22\%$ at
$\rho=0.34$ fm$^{-3}$.

\acknowledgments{The work is supported by the National Natural
Science Foundation of China (Grant No. 11175219), the
Major State Basic Research Developing Program of China (Grant No. 2013CB834405),
the Knowledge Innovation Project (Grant No. KJCX2-EW-N01) of
Chinese Academy of Sciences, the Chinese Academy of Sciences via a grant for
visiting senior international scientists (Grant No. 2009J2-26),
the Project of Knowledge Innovation Program (PKIP) of Chinese
Academy of Sciences (Grant No. KJCX2.YW.W10),
and the CAS/SAFEA International Partnership Program
for Creative Research Teams (Grant No. CXTD-J2005-1).}


\end{document}